\documentclass[notitlepage,showpacs,preprintnumbers,amsmath,amssymb,superscriptaddress,prl,twocolumn]{revtex4-1}
\pdfoutput=1
\usepackage{graphicx}
\usepackage[T1]{fontenc} 
\usepackage[utf8]{inputenc}
\usepackage[normalem]{ulem}

\newcommand{\mnu}{\ensuremath{\Sigma m_\nu}}
\newcommand{\Alens}{\ensuremath{A_{\rm{lens}}}}
\newcommand{\Neff}{\ensuremath{N_{\rm{eff}}}}
\newcommand{\lcdm}{$\Lambda$CDM}
\newcommand{\mcm}[1]{\ensuremath{\mathcal{M}_{#1}}}

\begin{document}
\title{Cosmology-marginalized approaches in Bayesian model comparison:\\
the neutrino mass as a case study}

\author{S.\ Gariazzo}\email{gariazzo@ific.uv.es}

\author{O.\ Mena}\email{omena@ific.uv.es}

\affiliation{Instituto de F\'isica Corpuscular (CSIC-Universitat de Val\`encia), Paterna (Valencia), Spain}

\begin{abstract}
We propose here a \emph{novel} method which singles out the \emph{a priori}
unavoidable dependence on the underlying cosmological model when
extracting parameter constraints, providing robust limits which only depend on the considered dataset.
Interestingly, when dealing with several possible cosmologies
and interpreting the Bayesian preference in terms of the Gaussian statistical evidence, the preferred model is much
less favored than when only two cases are compared.
As a working example, we apply our approach to the cosmological
neutrino mass bounds, which play a fundamental role
not only in establishing the contribution of relic neutrinos to the dark matter of the Universe,
but also in the planning of future experimental searches of the neutrino character and of the neutrino mass ordering.
\end{abstract}

\maketitle

{\it\textbf{Introduction---}} Bayesian parameter inference
has been extremely successful in cosmology and
astroparticle physics in the past two decades. This statistical
technique is more powerful and adequate than traditional tools when dealing
with large and complex data sets and with the impossibility to obtain
different realizations of the object to study, our universe.
In addition, the Bayesian
probability theory has also been extensively exploited for model
comparison purposes, offering not only the
possibility of predicting but also of optimizing the
most adequate theoretical frameworks to fit the cosmological
observations, see e.g.~\cite{Trotta:2008qt}.
However, despite the
major accomplishments achieved by Bayesian parameter
inference, both the role of parameterizations/priors and the possibility of different
fiducial cosmologies (or models) may led to divergent predictions.
The former have caused controversial arguments in the
literature, particularly when extracting cosmological bounds
on neutrino masses and on their ordering~\cite{Simpson:2017qvj,Schwetz:2017fey,Gariazzo:2018pei,Long:2017dru,Heavens:2018adv,Handley:2018gel}.

In this \textit{letter}, we shall focus on the potential that Bayesian model
comparison techniques offer for computing model-marginalized
cosmological parameter limits, avoiding the biases due to the
fiducial cosmology.
We propose here a simple method
to compute such solid and robust model-marginalized constraints.

In order to demonstrate the validity and robustness of this method,
we shall illustrate a particular case and consider the sum of the
neutrino mass $\mnu$ (see Refs.~\cite{Lesgourgues-Mangano-Miele-Pastor-2013,Lattanzi:2017ubx,deSalas:2018bym}
for its key signatures on cosmology).
Focusing exclusively on bounds from
Cosmic Microwave Background (CMB) measurements, the final analyses from the Planck
satellite set a $95\%$~CL limit of $\mnu <0.24$~eV~\cite{Aghanim:2018eyx}
after considering CMB temperature, polarization and lensing at all scales.
Late-time observations of the large scale structure in the universe by
means of the Baryon Acoustic Oscillation (BAO) method sharpen
the limit above, as they help enormously in removing the degeneracies
present in CMB data at the background level.
Once BAO information is
combined with Planck measurements, the limit is tightened to
$\mnu <0.12$~eV at $95\%$~CL~\cite{Aghanim:2018eyx}, or even down to
$\mnu <0.11$~eV when also considering Supernovae Ia luminosity distances.
One obvious question is: how reliable and stable the cosmological
neutrino mass limits quoted above are?

Even if not relying on the combination of potentially inconsistent data sets, for which the neutrino
mass bounds become tighter~\footnote{As an example, adding the prior on the
Hubble constant provided by Ref.~\cite{0004-637X-855-2-136} one would get $\mnu
<0.0970$~eV at $95\%$~CL~\cite{Aghanim:2018eyx}).}, all of the
aforementioned limits are based on the most economical \lcdm\ scenario,
which also leads to the tightest constraints on the neutrino mass.
Surely, the bounds on $\mnu$ change when
\textit{(a)} new physics is added in the neutrino sector (for instance, changing the
effective number of relativistic degrees of freedom, $\Neff$~\cite{Hamann:2007pi,Hamann:2010bk,Giusarma:2011ex,Hamann:2011ge,Giusarma:2012ph,RiemerSorensen:2012ve,Archidiacono:2013fha,DiValentino:2013qma,Archidiacono:2013lva}
or adding non-standard
interactions~\cite{Beacom:2004yd,Bell:2005dr,Hannestad:2004qu,Fardon:2003eh,Afshordi:2005ym,Brookfield:2005bz,Brookfield:2005td,Bjaelde:2007ki,Mota:2008nj,Ichiki:2008rh,Boehm:2012gr,Archidiacono:2013dua,Dvali:2016uhn,DiValentino:2017oaw}),
\textit{(b)} new physics appears in the early or late-time accelerating periods in the
universe~\cite{Hamann:2006pf,Joudaki:2012fx,Archidiacono:2013lva,dePutter:2014hza,DiValentino:2016ikp,Canac:2016smv,Gerbino:2016sgw,DiValentino:2016ikp,Gerbino:2016sgw,Huterer:2006mva,Baldi:2013iza,Hu:2014sea,Shim:2014uta,Barreira:2014ija,Bellomo:2016xhl,Peirone:2017vcq,Renk:2017rzu,Dirian:2017pwp,Gavela:2009cy,Reid:2009nq,Honorez:2009xt,LaVacca:2008mh,Guo:2018gyo}
 and/or, in general,
\textit{(c)} phenomenologically extended
scenarios are considered~\cite{DiValentino:2015ola}.
While one would naively expect that the neutrino mass limits within
these more general cosmologies will always be relaxed, this has been shown
to not to be the case for physical dark energy models~\cite{Vagnozzi:2018jhn,Choudhury:2018byy}, for which the neutrino
mass bounds get tighter than those obtained in the \lcdm\ framework.
It is therefore clear that one can artificially tune the cosmological neutrino mass
limits in an optimistic or in a pessimistic manner.

These \emph{a priori} harmless uncertainties translate into very serious dilemmas for neutrino particle physics
searches.
The near and far future neutrinoless double beta decay roadmap provides a very important example.
It seems therefore mandatory to build a method to extract model-independent
cosmological neutrino mass bounds.
It is among our major goals to
apply our novel model-marginalized method to $\mnu$ when studying a number of possible cosmological
scenarios, i.e.\ the minimal \lcdm\ universe with massive
neutrinos and its extensions.
Adopting Planck 2015 data
\footnote{Although the final data analyses have already been presented
by the Planck collaboration, the data and likelihood codes are not
public at the time of writing.}, the tightest bound we obtain within a \lcdm\
universe is $\mnu<0.23$~eV at $95\%$~CL, which relaxes to
$\mnu<0.35$~eV when the uncertainty on the cosmological model is
taken into account using our model-marginalization method.

At the same time, we can use Bayesian tools in order to compare
the models we are studying and obtain which one is preferred by data. 
Noticeably, even if the best scenario is strongly favoured over its
competitors when comparing pairs of models with a Bayes factor
analysis, its global statistical evidence falls abruptly
when all the models are considered simultaneously,
making this preferred model less likely.
In the scenarios explored here,
this will imply that the weak-to-moderate Bayesian preference for the minimal
\lcdm+\mnu\ model, which arises when it is compared with each of its
extensions individually, will not correspond to a global $1\sigma$ level
strength when considering the entire ensemble of extended scenarios.

\medskip
{\it\textbf{Bayesian statistics---}} The Bayes theorem, which represents the foundation of Bayesian statistics, reads:
\begin{equation}\label{eq:bayestheorem}
p(\theta|d,\mcm{i})
=
\frac{\pi(\theta|\mcm{i}) \mathcal{L}(\theta)}{Z_i}\,,
\end{equation}
where
$\pi(\theta|\mcm{i})$ and $p(\theta|d,\mcm{i})$ are the prior and posterior probabilities
for the parameters $\theta$ within a model \mcm{i},
$\mathcal{L}(\theta)$ is the likelihood as a function of the parameters $\theta$, given the data $d$ and the model \mcm{i},
and
$Z_i=\int d\theta\,\pi(\theta|\mcm{i})\,\mathcal{L}(\theta)$
is the Bayesian evidence of \mcm{i}~\cite{Trotta:2008qt}.
The Bayes theorem can also be written in a slightly different form to obtain the model posterior probability~\cite{Handley:2015aa}:
\begin{equation}\label{eq:bayestheoremmodel}
p_i
\equiv
p(\mcm{i}|d)
=\frac{\pi_i Z_i}{\sum_j \pi_j Z_j}~,
\end{equation}
where $\pi_i\equiv\pi(\mcm{i})$ refers to the model prior probability.
In the Bayesian model comparison framework,
the so-called \emph{Bayes factor} provides a measure of
whether the data have increased or decreased the odds of model
\mcm{i} relative to a second model \mcm{j}:
\begin{equation}\label{eq:bayesfactor}
B_{ij}
=
Z_i/Z_j\,.
\end{equation}
The Bayes factor enters the definition of the posterior probability ratio between two models,
which indicates how much one of the two is preferred over the other, after using the information provided by data:
\begin{equation}\label{eq:modelposteriors}
\frac{p_i}{p_j}
=
B_{ij}
\frac{\pi_i}{\pi_j}\,.
\end{equation}
If the two models are equivalent according to our initial knowledge,
i.e.\ the model priors are the same, the final preference driven by data is
determined by the Bayes factor.
In terms of posterior odds,
the preference for the favored model is $B_{ij}:1$, if \mcm{i} is preferred over \mcm{j}.
Adopting the
commonly exploited Jeffreys' scale~\cite{Jeffreys:1961a}, the strength
of the posterior odds can be ranked as
inconclusive ($|\ln B_{ij}|< 1$),
weak ($1<|\ln B_{ij}|<2.5$),
moderate ($2.5<|\ln B_{ij}|<5$), or
strong ($|\ln B_{ij}|>5$).
Very importantly, this arises from the fact that when comparing two mutually exclusive models,
the mentioned ranks correspond roughly to what is usually indicated as
$\lesssim1\sigma$ (inconclusive) to $\gtrsim3\sigma$ (strong) level
when considering a Gaussian variable.

Using Eq.~\eqref{eq:bayestheoremmodel}, and selecting one among the available models,
labelled \mcm{0} without loss of generality, one can write, provided
all priors are identical for all models:
\begin{equation}\label{eq:posterior_model0}
p_0
=\frac{Z_0}{\sum_i Z_i}
=
\left(
\sum_i^N B_{i0}
\right)^{-1}
\,,
\end{equation}
where we have used the definition of the Bayes factor.
Notice that the
posterior probability of the selected model \mcm{0} depends on the
Bayes factors with respect to all the possible models.
For each data combination we will choose \mcm{0} to be the preferred model.
In practice, this is the one that has more influence on the model-marginalized posterior.
Since the model \mcm{0} is the preferred one, we will always have
$B_{i0}=Z_i/Z_0 < 1$ (or $\ln B_{i0}<0$) for $i\neq0$.

Assuming that \textit{(i)} more than two models are possible; and
\textit{(ii)} all the models have the same prior probabilities, then
Eq.~\eqref{eq:posterior_model0} implies that the posterior probability of the preferred model is
smaller than what the single Bayes factors would suggest
in a one-to-one comparison.
For example, if $N=8$ and all the Bayes factors are $|\ln B_{i0}|\simeq 5$ for $i\neq0$,
thus indicating apparently strong results according to the usually adopted Jeffreys' scale,
the posterior probability of \mcm{0} is $p_0\simeq0.955$,
which would indicate a mild $2\sigma$ significance for a Gaussian measure.
In the same way, having $N=7$ and $|\ln B_{i0}|\simeq2.5$ for $i\neq0$,
which usually indicates a weak preference, would give $p_0\simeq
0.67$, which would correspond to less than $1\sigma$ preference for \mcm{0}.

The tools of model comparison also allow us to compute a model-marginalized posterior distribution
for the parameter $\theta$, taking into account the posterior probability of each model \mcm{i}
resulting from the data $d$ \cite{Trotta:2008qt}:
\begin{equation}\label{eq:modelmarginalize}
p(\theta|d)
=
\sum_i^N
p(\theta|d, \mathcal{M}_i)\,
p_i\,,
\end{equation}
where the posterior probabilities of $\theta$ within each model \mcm{i}
are weighted according to the model posterior probabilities $p_i$.
These can be written using Eq.~\eqref{eq:bayestheoremmodel}
to obtain the fundamental formula
\begin{equation}\label{eq:modelmarginalizefinal}
p(\theta|d)
=
\left.
\sum_i^N
p(\theta|d, \mathcal{M}_i) Z_i
\right/
\sum_j^N Z_j
\,.
\end{equation}
This is the expression that we will use to obtain model-marginalized limits in the following,
under the assumption that all the models have the same priors.

Some final comments are due.
To obtain the most robust model-marginalized estimate one should in principle consider
the largest number of possible models.
In the cosmological context, these should include the \lcdm\ and all its possible extensions,
plus scenarios with any possible modified gravity paradigm and their extensions:
this is clearly computationally impossible.
From an Occam's razor perspective, however, the models with an unnecessarily large number of parameters
will be generally penalized by the Bayesian evidence calculation~%
\footnote{This is true as long as the additional parameters are constrained by the data,
as no penalty applies to unconstrained parameters.
The selection of models considered in the analysis should take into account equal priors only
for reasonable extensions of the simplest model.},
so that their final weight in Eq.~\eqref{eq:modelmarginalizefinal} will be negligible,
while most of the contribution will be given by the most economical models that better fit the data.
While our method allows to marginalize over the freedom related to different models or additional parameters,
since it is based on the comparison of Bayesian evidences obtained in the different models,
it still has a residual dependence on the shape and the width of the adopted priors.

\medskip
{\it \textbf{Cosmological data analyses---}}
The data we shall exploit to derive model-marginalized
constraints from cosmological observations include measurements of
the CMB angular power spectrum and of the BAO signature in the
matter power spectrum.
Awaiting for the final release from
the Planck collaboration, we use here their 2015 data
release~\cite{Adam:2015rua,Ade:2015xua}.
We consider two possibilities:
\textit{a)} both temperature
and low-$\ell$ polarization (CMB), or
\textit{b)} temperature and polarization at all multipoles (CMB+pol).
In both cases we also include the Planck CMB lensing determination (lens)~\cite{Ade:2015zua}.
BAO geometrical information from the \texttt{SDSS BOSS} DR11
\cite{Anderson:2013zyy}, the \texttt{6DF}~\cite{Beutler:2011hx} and
the \texttt{SDSS DR7 MGS}~\cite{Ross:2014qpa} surveys complements
the data sets used in our numerical analyses.
We are aware that this combination may not provide the strongest
cosmological constraints.
However, it is not our main goal here to outperform
the current cosmological constraints, but to exemplify the novel model-marginalized
approach here proposed.
After the Planck final public release, our
method will be applied to an extended set of cases with respect to those considered here.

In our numerical calculations we use
the Boltzmann solver \texttt{CAMB} \cite{Lewis:1999bs}
together with \texttt{CosmoMC} \cite{Lewis:2002ah},
with \texttt{PolyChord}
\cite{Handley:2015fda,Handley:2015aa} (version \textbf{1.9}) as the algorithm devoted to
extract the Bayesian evidences.

In our demonstrative analysis,
we restrict our set of models to the simplest \lcdm\ model with freely varying neutrino masses
and some of its one-parameter extensions.
In particular, we consider the \lcdm+\mnu, \lcdm+\mnu+\Alens, \lcdm+\mnu+\Neff\ and \lcdm+\mnu+$w$ models,
as discussed more in detail in the next paragraphs.
In the numerical calculations, all the parameters that are shared among the different models are sampled adopting
the same linear priors as in the default \texttt{PolyChord} settings,
except for the sum of the neutrino masses which is varied in the range $[0.06,\, 5]$~eV.
For the additional parameters we adopt linear priors in the following ranges:
\Alens\ varies in $[0,\, 5]$, \Neff\ in $[1,\,5]$ and $w$ in $[-3,\,0]$.

\medskip
{\it \textbf{Results: the neutrino mass as a case study---}}
Table~\ref{tab:limits_all} summarizes the results from our novel method
applied to a particular physics case that is usually constrained by
cosmological observations: the sum of the neutrino masses $\mnu$.
As aforementioned, a robust model-marginalized limit
on $\mnu$ is absolutely required, as it is crucial for a number of issues.
In particular, it is a very important input when deciding the
experimental strategy for neutrino character (Dirac versus
Majorana) searches.
We show such model-marginalized limit in the second-to-last row of
Tab.~\ref{tab:limits_all}, for the two data combinations considered here.

\begin{table*}[t]
\centering
\begin{tabular}{l||c|c||c|c}
&\multicolumn{2}{c||}{~~CMB+lens+BAO~~}
&\multicolumn{2}{c} {~~CMB+pol+lens+BAO~~} \\
model
& ~$\ln B_{i0}$~ & \mnu\ [eV]
& ~$\ln B_{i0}$~ & \mnu\ [eV] \\
\hline
base=$\Lambda$CDM+$\Sigma m_\nu$ & $     0.0$ & $        <  0.28$ & $     0.0$ & $        <  0.23$ \\
             base+$A_{\rm lens}$ & $    -2.6$ & $        <  0.38$ & $    -2.4$ & $        <  0.29$ \\
              base+$N_{\rm eff}$ & $    -1.5$ & $        <  0.37$ & $    -2.3$ & $        <  0.25$ \\
                        base+$w$ & $    -1.4$ & $        <  0.42$ & $    -0.1$ & $        <  0.42$ \\
\hline
                    marginalized & $       -$ & $        <  0.33$ & $       -$ & $        <  0.35$ \\
\hline
$p_0$ &\multicolumn{2}{c||}{  0.65} & \multicolumn{2}{c}{  0.48} \\
\end{tabular}
\caption{\label{tab:limits_all}
$95\%$~CL upper limits on $\mnu$ and Bayes factors in the different cosmological scenarios.
Results are obtained either adopting Planck 2015 CMB temperature,
low-$\ell$ polarization and lensing data~\cite{Adam:2015rua,Ade:2015xua,Ade:2015zua}
plus BAO
measurements~\cite{Anderson:2013zyy,Beutler:2011hx,Ross:2014qpa}
(second and third columns) or
the same data combination plus high-multipole CMB
polarization measurements from the Planck 2015 data release (fourth
and fifth columns).
The different rows depict the bounds in different
extensions of the \lcdm\ model, while the last two rows illustrate,
respectively, the
model-marginalized $95\%$~CL limit obtained via Eq.~\eqref{eq:modelmarginalizefinal}
and the posterior probability of the example model \mcm{0}
(the preferred one, that is always the \lcdm+\mnu\ scenario), see Eq.~\eqref{eq:posterior_model0}.
}
\end{table*}

In order to compute the model-marginalized result
we consider, together with the simplest \lcdm\ model with freely varying neutrino masses,
some cosmological scenarios which are usually explored in the
literature, see e.g.\ Planck 2015 data analyses~\cite{Ade:2015xua}.
These models contain extra parameters which are either partially or
significantly degenerate with the neutrino mass.
For instance, \mnu\ has a correlation, among others,
with the phenomenological parameter $\Alens$,
which rescales the lensing amplitude in the CMB spectra.
Since current CMB constraints on the neutrino mass are
mostly due to the reduction in the lensing potential induced by a
larger neutrino mass, there is a degeneracy between $\mnu$ and
$\Alens$: a value $\Alens <1 \ (>1)$ would allow for a lower (higher)
value of $\mnu$.
Notice from the results depicted in
Tab.~\ref{tab:limits_all} that the neutrino mass bounds, in
absence of high-multipole polarization, are worsened
when the $\Alens$ parameter is allowed to vary.
When going from a \lcdm\ to a \lcdm+$\Alens$ scenario the $95\%$~CL limit changes
from $\mnu <0.28$~eV to $\mnu <0.38$~eV~\footnote{As a comparison, the Planck collaboration 2015
data analyses quote $\mnu <0.25$~eV and $\mnu <0.41$~eV for these two
cases, clearly stating the very good consistency of the results here
obtained.}.
Another parameter potentially degenerate with
$\mnu$ is the number of relativistic degrees of freedom
$\Neff$, albeit the latest Planck analyses have shown that data
are able to disentangle between the different physical effects
induced by $\mnu$ and $\Neff$~\cite{Aghanim:2018eyx} on temperature and polarization anisotropies.
While the $95\%$~CL limit
without high-$\ell$ polarization is $\mnu <0.37$~eV,
information from high multipoles brings the neutrino mass constraint extremely
close to the bound obtained within the \lcdm\ model~\footnote{The 2015
Planck results provide a $95\%$~CL upper limit of $\mnu
<0.32$~eV ($\mnu <0.22$~eV) without (with) the addition of
high-$\ell$ polarization measurements in the \lcdm\ +
$\Neff$ model.}.
Finally, a freely-varying constant dark energy equation of state $w$
can also affect the bounds on $\mnu$.
If $w$ is allowed to vary, the matter energy
density can take very high values, compensating for the suppression
induced in large scale structure due to an increased value of $\mnu$,
and therefore these two parameters will be correlated in a significant
way.
As a result, the limit is relaxed to $\mnu <0.42$~eV at $95\%$~CL,
both without and with high-multipole polarization data, see
Tab.~\ref{tab:limits_all}.

\begin{figure}[t]
\centering
\includegraphics[width=\columnwidth]{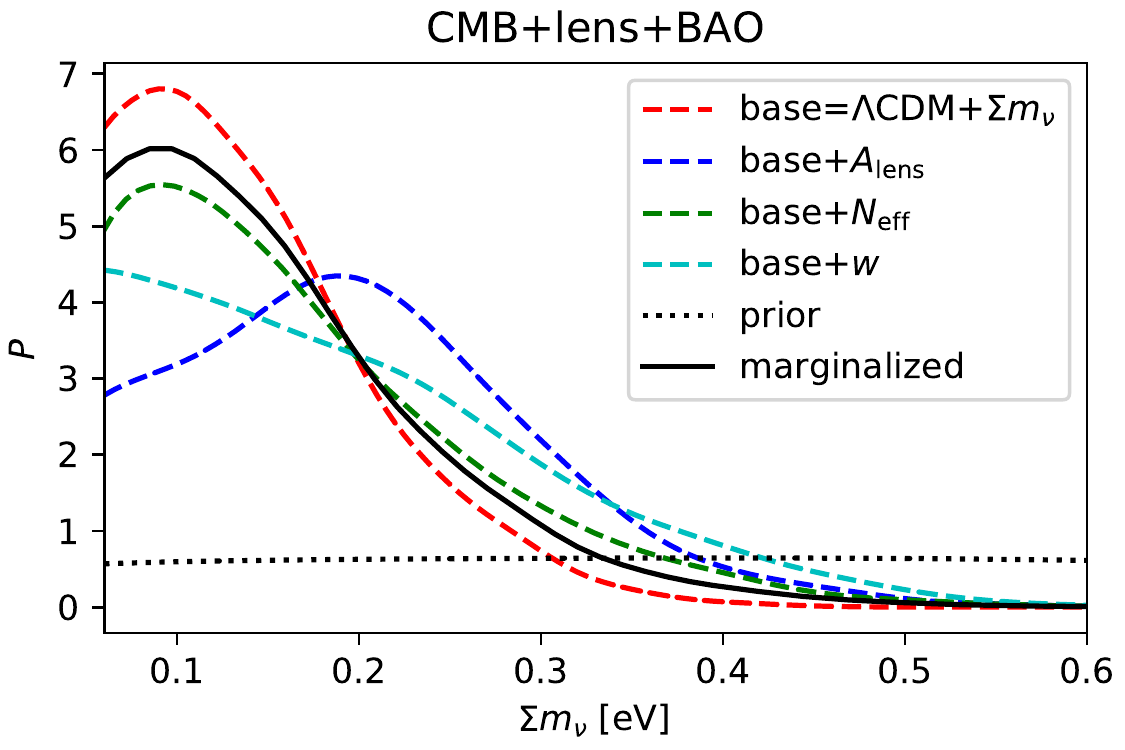}
\caption{\label{fig:cmblb_res}
One-dimensional posterior probabilities for \mnu\ for different cosmological models,
arising from Planck 2015 CMB temperature,
low-$\ell$ polarization and lensing data~\cite{Adam:2015rua,Ade:2015xua,Ade:2015zua}
plus BAO measurements~\cite{Anderson:2013zyy,Beutler:2011hx,Ross:2014qpa}.
We also depict the model-marginalized bound
obtained using Eq.~\eqref{eq:modelmarginalizefinal} and the prior
sampled on $\mnu$, see text for details.}
\end{figure}

\begin{figure}[t]
\centering
\includegraphics[width=\columnwidth]{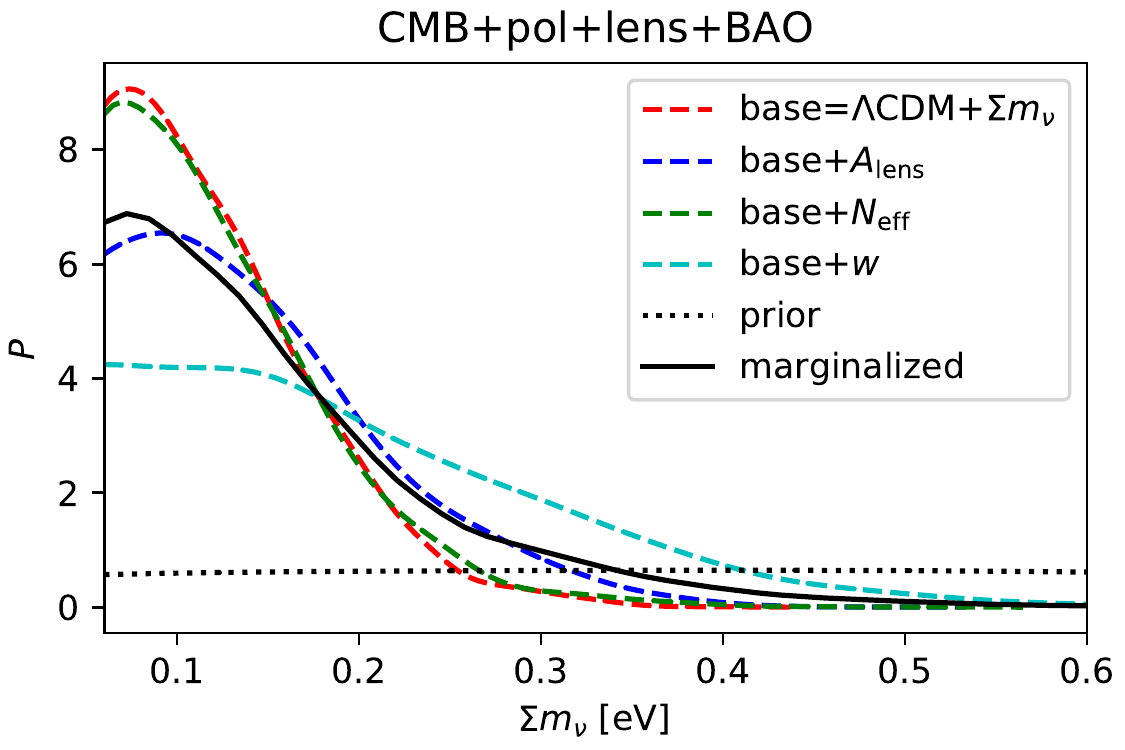}
\caption{\label{fig:cmblbp_res}
As Fig.~\ref{fig:cmblb_res}, but
including also CMB high multipole polarization measurements from the Planck 2015 data
release~\cite{Adam:2015rua,Ade:2015xua}.}
\end{figure}

From all the limits above and using the Bayes factors also listed in
Tab.~\ref{tab:limits_all}, by means of Eq.~\eqref{eq:modelmarginalizefinal}
it is possible to obtain the marginalized limits on $\mnu$ shown in the
second-to-last row of Tab.~\ref{tab:limits_all}.
Notice that the $95\%$~CL upper
limits obtained within the most economical \lcdm\ picture are
significantly relaxed (they are increased up to $50\%$)
when considering extended scenarios.
For a visual comparison of the one-dimensional posterior probabilities of \mnu\
in the various models considered here and of the model-marginalized one,
we provide Figures \ref{fig:cmblb_res} and \ref{fig:cmblbp_res}, where
we also show the sampled prior distribution~\footnote{
While, in principle, the prior sampled by \texttt{CosmoMC} can be non-trivial,
the distribution turns out to be flat in the parameter region of
interest here, making our results strongly independent on the prior choice.}
Notice that the method following Eq.~\eqref{eq:modelmarginalizefinal}
allows a proper weighing of the information from each model, building a robust estimate for the
neutrino mass that can be used as an input in neutrino particle physics.
The possible applications of the method, however, are significantly wider than what explored here.

The last row of Tab.~\ref{tab:limits_all} shows the posterior
probabilities $p_0$ for the \emph{example} model \mcm{0}, computed from Eq.~\eqref{eq:posterior_model0}.
\mcm{0} is chosen to be the preferred one by
each of the two data combinations, and it turns out to be
the minimal \lcdm\ scenario with free neutrino masses in both cases.
The posterior probability, which depends on the Bayesian evidences of various models,
is shown in the second and fourth columns for the two possible data
combinations.
Here one should clarify an important aspect of Bayesian model comparison.
While the Bayes factors with respect to the extended models,
if considered separately,
indicate a weak-to-moderate~\footnote{One exception is represented by the \lcdm+$w$ model
when full polarization data are included, which is basically equivalent to the \lcdm\ model.
All the Bayes factors we obtain from our numerical calculation, however, carry an uncertainty of $\sim0.3$.}
Bayesian preference for the \lcdm\ model accordingly to
the Jeffreys' scale~\cite{Jeffreys:1961a}, and therefore individually corresponding to a $1.1-2.7\sigma$ probability
(in Gaussian terms) in favor of the \lcdm\ framework,
it is clear from Eq.~\eqref{eq:posterior_model0}
that such naive expectations are no longer true when more than one
model is accesible.
The values of the posterior probabilities for the
\emph{example} \lcdm\ model never reach the $1\sigma$ level
strength in terms of a Gaussian variable.
Based on these results, therefore, it is possible to say
that the \lcdm\ model, despite being more likely than its extensions,
is not strongly preferred by data.
This is a crucial result of our analyses, with strong implications in many other
early universe fundamental physics searches,
as for example in the case of the inflationary landscape,
where many models arise and are usually ranked by means of Bayesian comparison techniques,
see e.g.\ Ref.~\cite{Ade:2015lrj}.

\medskip
{\it \textbf{Discussion---}} Bayesian model comparison provides a robust machinery to
compute model marginalized limits.
We have proposed a method which allows to minimize the uncertainty related to multiple model choices on
the determination of parameter constraints.
We have applied our novel method to the neutrino mass case,
exploiting current publicly available cosmological data.
We show that the limits on the neutrino
masses can significantly change when one realizes that present
measurements are not able to unambiguously tell us the cosmological model that
nature has chosen.
The statistical \emph{Gaussian} preference for the favoured model, indeed,
always becomes inconclusive when there are a number
of other possible models, even if equally disfavored by observations.
An updated and extended analysis using the proposed method
will come after the release of Planck 2018 likelihoods.

\bigskip
{\it \textbf{Acknowledgments---}}
We thank J.\ Lesgourgues for his presentation at Neutrino 2018, which inspired the idea of this paper,
and M.\ Lattanzi and S.~Pastor for useful comments on the draft and
enlightening discussions.
Work supported by the Spanish grants
FPA2015-68783-REDT, 
FPA2017-90566-REDC (Red Consolider MultiDark),
FPA2017-85216-P, 
FPA2017-85985-P 
and
SEV-2014-0398 
(AEI/FEDER, UE, MINECO),
and PROMETEOII/2014/050, 
PROMETEOII/2018/165 
and GV2016-142 (Generalitat Valenciana).
SG receives support from the European Union's Horizon 2020 research and innovation programme under the Marie Sk{\l}odowska-Curie individual grant agreement No.\ 796941.
OM is also supported by the European Union's Horizon 2020 research and innovation program under the Marie
Sk\l odowska-Curie grant agreements No.\ 690575 and 674896.

\bibliography{main}

\end{document}